\documentclass[journal]{IEEEtran}

\usepackage{cite}
\usepackage{amsmath,amssymb,amsfonts,bm}
\usepackage{algorithm}
\usepackage{algorithmic}
\usepackage{graphicx}
\usepackage{epstopdf}
\usepackage{booktabs}
\usepackage{makecell}
\usepackage{multirow}
\usepackage{textcomp}
\usepackage{xcolor}
\usepackage[caption=false,font=normalsize,labelfont=sf,textfont=sf]{subfig}
\usepackage[colorlinks  = true,
            linkcolor   = black,
            urlcolor    = magenta,
            anchorcolor = blue,
            bookmarks   = false
            ]{hyperref}

\def\BibTeX{{\rm B\kern-.05em{\sc i\kern-.025em b}\kern-.08em
    T\kern-.1667em\lower.7ex\hbox{E}\kern-.125emX}}

\begin{document}

\title{Towards Efficient Subarray Hybrid Beamforming: Attention Network-based Practical Feedback in FDD Massive MU-MIMO Systems}  

\author{

Zhilin Lu, Xudong Zhang, Rui Zeng and Jintao Wang,~\IEEEmembership{Senior Member,~IEEE}

\thanks{

The authors are with the Department of Electronic Engineering, Tsinghua University, and Beijing National Research Center for Information Science and Technology (BNRist), Beijing 100084, China. (e-mail: luzl18@mails.tsinghua.edu.cn, zxd22@mails.tsinghua.edu.cn, zengr21@mails.tsinghua.edu.cn, wangjintao@tsinghua.edu.cn).

The key results can be reproduced with the following github link: \textnormal{\href{https://github.com/Kylin9511/EFBAttnNet}{https://github.com/Kylin9511/EFBAttnNet}}.

}
}

{}

\maketitle

\begin{abstract}
Channel state information (CSI) feedback is necessary for the frequency division duplexing (FDD) multiple input multiple output (MIMO) systems due to the channel non-reciprocity.
With the help of deep learning, many works have succeeded in rebuilding the compressed ideal CSI for massive MIMO.
However, simple CSI reconstruction is of limited practicality since the channel estimation and the targeted beamforming design are not considered.
In this paper, a jointly optimized network is introduced for channel estimation and feedback so that a spectral-efficient beamformer can be learned.
Moreover, the deployment-friendly subarray hybrid beamforming architecture is applied and a practical lightweight end-to-end network is specially designed.
Experiments show that the proposed network is over 10 times lighter at the resource-sensitive user equipment compared with the previous state-of-the-art method with only a minor performance loss.
\end{abstract}


\section{Introduction}

\IEEEPARstart{M}{assive} multiple input multiple output (MIMO) is widely regarded as an important technique in future wireless communication systems.
The base station (BS) requires the downlink channel state information (CSI) for beamforming to achieve better spectrum efficiency \cite{Molisch2017hybrid}.
In frequency division duplexing (FDD) modes, the downlink CSI needs to be estimated at the user equipment (UE) and fed back to the BS due to the channel non-reciprocity.
However, the overhead of direct feedback is unacceptably large due to the large antenna scale in massive MIMO systems.

Deep learning (DL) aided CSI compressed feedback has drawn wide attention since the CsiNet\cite{wen2018deep} showed great superiority over the traditional compressed sensing algorithms.
Nevertheless, many existing works like CRNet\cite{lu2020multiresolution} assume the ideal channel estimation and simply aim at better CSI reconstruction, which can not guarantee a higher sum rate of beamforming.
Therefore, both channel estimation and beamforming should be considered for a practical feedback.

For a traditional block-based communication system, it is hard to optimize the adjacent blocks like CSI feedback and beamforming together.
On the contrary, it is natural to concatenate the DL networks for adjacent blocks and jointly optimize the whole pipeline in an end-to-end manner.
For instance, an autoencoder for joint channel estimation and feedback is built to improve the channel reconstruction accuracy in \cite{Ma2021model}, while the feedback and beamforming are optimized together for a higher beamforming gain in \cite{guo2021deep-FB-BF}.

Furthermore, some works try to jointly optimize the channel \textit{Estimation}, \textit{Feedback}, and \textit{Beamforming} (EFB) to improve the beamforming sum rate systematically.
A digital beamforming aided EFB pipeline is realized in \cite{Sohrabi2021deep} for multiuser MIMO (MU-MIMO) systems. Further, the EFB networks under hybrid beamforming architecture are proposed in \cite{Gao2022data} and \cite{Wu2022deep}.
However, existing works on the EFB joint optimization are not friendly enough to the deployment since they only consider the fully-connected beamforming structure and the designed networks are rather heavy.

In this paper, we focus on improving the practicality of the DL-based EFB joint optimization methods. The main contributions are summarized as follows.
\begin{itemize}
    \item The sub-connected hybrid beamforming structure is first introduced to the EFB joint optimization task. The energy efficiency is improved by reducing the number of required phase shifters (PSs).
    \item A low complexity EFB network is designed to improve the practicality. Compared with the previous state-of-the-art (SOTA) method, the proposed EFBAttnNet provides a 10 times encoder complexity reduction at the resource-sensitive UE with only a small performance loss.
\end{itemize}

The rest of the paper is organized as follows. Section \ref{SectionSystemModel} introduces the system model and section \ref{SectionProblemFormulation} explains the EFB problem formulation. The lightweight EFBAttnNet is proposed in section \ref{SectionEFBAttnNet} and the experimental results are analyzed in section \ref{SectionNumericalResults}. Finally, the conclusion is drawn in section \ref{SectionConclusion}.

\section{System Model} \label{SectionSystemModel}

We consider an FDD massive MU-MIMO system with one BS and $K$ independent UEs. The BS has $N_t$ antennas and $K$ radio frequency (RF) chains while each UE has one antenna. The received signal at the $k^{th}$ user can be derived as follows:

\begin{equation} \label{EquationReceivedSignal}
    y_k = \mathbf{h}_k^H\mathbf{w}_ks_k + \sum_{i=1,i\neq k}^{K}\mathbf{h}_k^H\mathbf{w}_is_i + z_k,
\end{equation}
where $\mathbf{h}_k\in\mathbb{C}^{N_t\times 1}$ and $\mathbf{w}_k\in\mathbb{C}^{N_t\times 1}$ are the downlink channel vector and the equivalent beamforming vector for the $k^{\text{th}}$ user, respectively. $s_k\in\mathbb{C}$ denotes the symbol sent to the $k^{\text{th}}$ user while $z_k \sim \mathcal{CN}\left(0,\sigma^2\right)$ stands for the additive white Gaussian noise at the receiver.

\begin{figure}[!t]
    \centering
    \includegraphics[width=\linewidth]{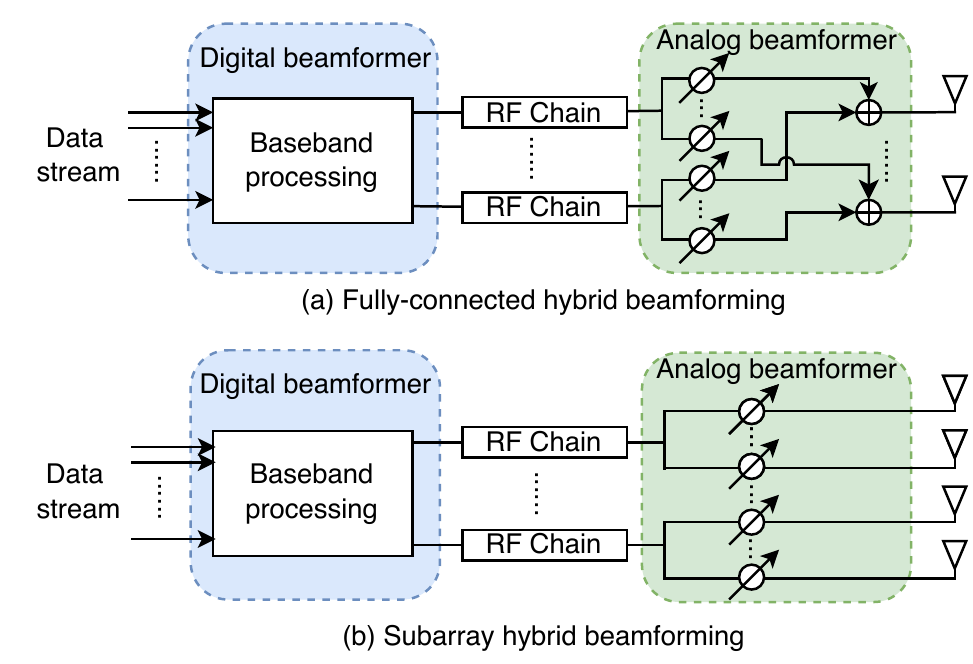}
    \caption{Architecture comparison between fully connected hybrid beamforming and subarray hybrid beamforming.}
    \label{ImageSubarray}
\end{figure}

Previous works on the EFB joint optimization with hybrid beamforming focus on the fully-connected architecture as shown in Fig. \ref{ImageSubarray} (a). In order to improve the energy efficiency, we first introduce the subarray beamforming structure to the EFB pipeline. As shown in Fig. \ref{ImageSubarray} (b), $N_t$ antennas are evenly distributed to $K$ RF chains. Each antenna is connected to only one RF chain instead of all RF chains so that the required number of phase shifters (PSs) is largely reduced. Therefore, the equivalent beamformer $\mathbf{w}_k$ follows the paradigm below.

\begin{equation} \label{EquationMergeBeamformer}
    \mathbf{w}_k = \mathbf{A}\mathbf{C}\mathbf{d}_k,
\end{equation}
where the analog beamformer $\mathbf{A} \triangleq \text{diag}\left(e^{j\boldsymbol{\theta}_\mathbf{A}}\right)\in\mathbb{C}^{N_t\times N_t}$ is a constant modulus diagonal matrix and $\boldsymbol{\theta}_\mathbf{A} \in \mathbb{R}^{N_t\times 1}$ is the phase vector. Meanwhile, each user has its own digital beamformer $\mathbf{d}_k \in \mathbb{C}^{K\times 1}$ and we denote the overall digital beamformer as $\mathbf{D}\triangleq \left[\mathbf{d}_1, \dots, \mathbf{d}_K\right]\in\mathbb{C}^{K\times K}$. The antenna connector $\mathbf{C}$ is a block diagonal matrix that links each antenna to one RF chain so that the subarray architecture is guaranteed.

\begin{equation} \label{EquationC}
    \mathbf{C} = \mathbf{I}_K \otimes \mathbf{1}_{N_m} \in\{0,1\}^{N_t\times K},
\end{equation}
where $\mathbf{I}_K \in \mathbb{R}^{K\times K}$ is an identity matrix and $\mathbf{1}_{N_m} \in \mathbb{R}^{N_m\times 1}$ is an all-one vector with $N_m=N_t/K$ representing the number of antennas connected to a single RF chain.

As for the channel $\mathbf{h}_k$, the influential clustered Saleh-Valenzuela (SV) model is adopted as equation (\ref{EquationSV}) shows.

\begin{equation} \label{EquationSV}
    \mathbf{h}_k = \frac{1}{L_\text{path}}\sum_{p=1}^{L_\text{path}}\alpha_{p,k}\mathbf{a}_t(\phi_{p,k}),
\end{equation}
where the $\alpha_{p,k}$ and the $\phi_{p,k}$ are the complex gain and the angle of departure (AoD) of the $k^{\text{th}}$ user at the $p^{\text{th}}$ path, respectively. The uniform linear array (ULA) model is applied, therefore the transmitter response vector $\mathbf{a}_t$ is set as:

\begin{equation}
    \mathbf{a}_t(\phi) = \left[ 1, e^{j\frac{2\pi}{\lambda}d\sin(\phi)}, \dots, e^{j\frac{2\pi}{\lambda}d(N_t-1)\sin(\phi)} \right]^T
\end{equation}

\section{The DL-based EFB Pipeline} \label{SectionProblemFormulation}

\begin{figure}[!b]
    \centering
    \includegraphics[width=\linewidth]{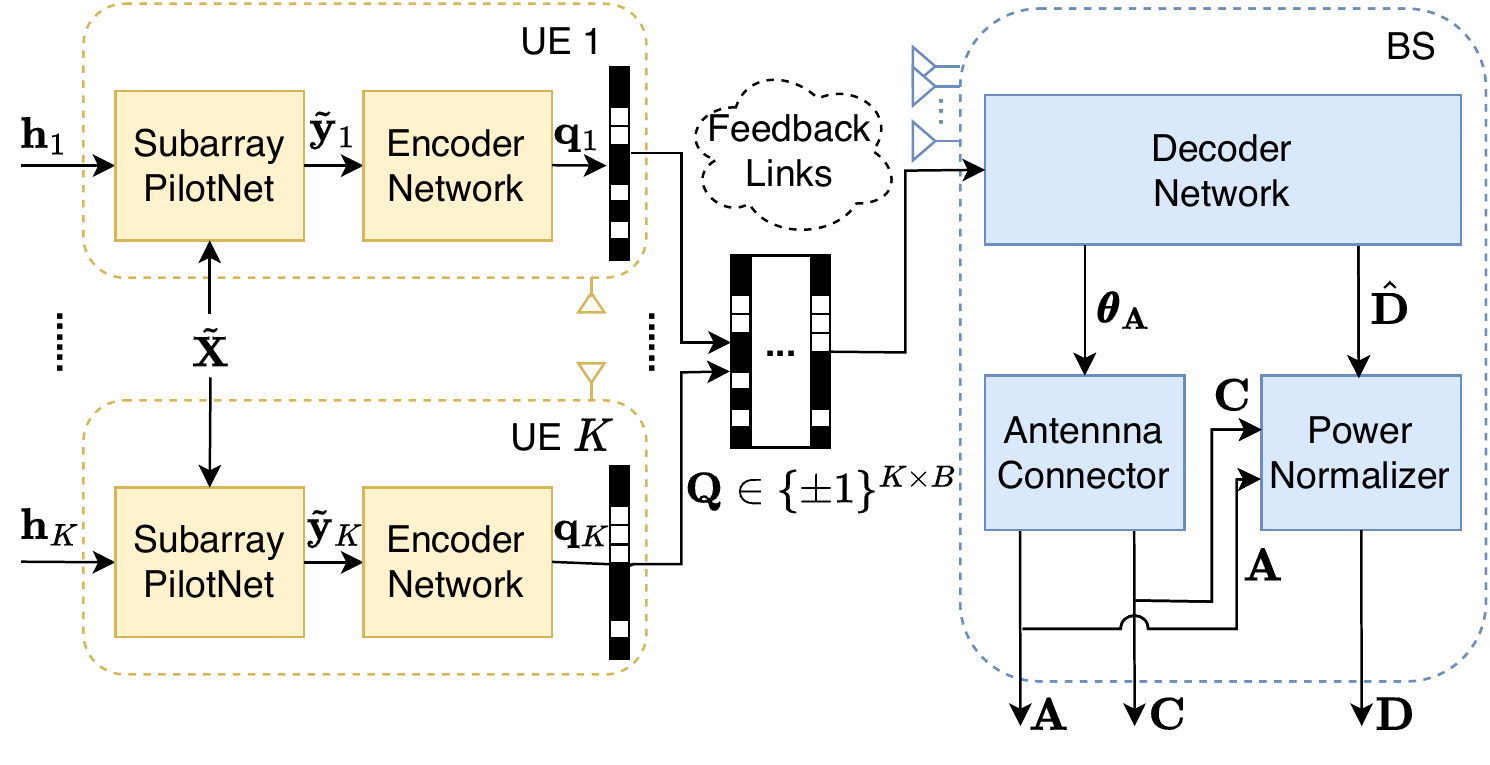}
    \caption{The DL-based EFB pipeline that co-designs the channel estimation, feedback, and beamforming with network joint optimization.}
    \label{ImageSystemModel}
\end{figure}

The DL-based EFB pipeline starts with the downlink pilot training where a fully-connected (FC) layer without bias is learned. Based on the subarray hybrid beamforming architecture, the weight $\boldsymbol{\theta}_\mathbf{\tilde{X}} \in \mathbb{R}^{N_t\times L}$ of the FC layer is used as the phases of the $L$ different pilots with the normalized amplitude.

\begin{equation}
    \mathbf{\tilde{X}} = \sqrt{P/N_t}e^{j\boldsymbol{\theta}_\mathbf{\tilde{X}}} = \sqrt{P/N_t}\left(\cos(\boldsymbol{\theta}_\mathbf{\tilde{X}}) + j\sin(\boldsymbol{\theta}_\mathbf{\tilde{X}})\right),
\end{equation}
where $P$ is the transmitting power. As shown in Fig. \ref{ImageSystemModel}, all users share the same pilots $\mathbf{\tilde{X}}$. The downlink pilot transmission realized by the SubarrayPilotNet can be described as:

\begin{equation}
    \mathbf{\tilde{y}}_k = \mathbf{h}_k^H\mathbf{\tilde{X}} + \mathbf{\tilde{z}}_k,
\end{equation}
where $\mathbf{\tilde{z}}_k\sim\mathcal{CN}(0, \sigma^2\mathbf{I}_L)$ is the additive white Gaussian noise.

For the encoder at the UEs, the channel estimation and the CSI compression are done together implicitly. As depicted in Fig. \ref{ImageSystemModel}, the feedback bitstream $\mathbf{q}_k \in \{\pm1\}^{B\times 1}$ is directly learned from $\tilde{\mathbf{y}}_k$.

\begin{equation}
    \mathbf{q}_k = \mathcal{E}(\mathbf{\tilde{y}}_k, \Theta_{\mathcal{E}});\;\; k\in\{1, \dots, K\},
\end{equation}
where $\mathcal{E}(\cdot)$ stands for the EFB encoder network and $\Theta_{\mathcal{E}}$ is the learnable parameters of the encoder.

For the decoder at the BS, the input $\mathbf{Q}=\left[\mathbf{q}_1, \dots, \mathbf{q}_K\right]$ gathers the bitstreams fed from the users. The phase vector $\boldsymbol{\theta}_\mathbf{A} \in \mathbb{R}^{N_t\times 1}$ and the raw digital beamformer $\hat{\mathbf{D}} \in \mathbb{C}^{K\times K}$ are set as the output of the decoder.

\begin{equation}
    \left(\boldsymbol{\theta}_\mathbf{A}, \hat{\mathbf{D}}\right) = \mathcal{D}(\mathbf{Q}, \Theta_{\mathcal{D}}),
\end{equation}

where $\mathcal{D}(\cdot)$ stands for the EFB decoder network with parameters $\Theta_{\mathcal{D}}$. Subsequently, a power normalization is carried out to meet the power constraint $\left\Vert \mathbf{A}\mathbf{C}\mathbf{D} \right\Vert^2_F\le P$ as follows.
\begin{equation}
    \mathbf{D} = \frac{\sqrt{P}\hat{\mathbf{D}}}{\left\Vert \mathbf{A}\mathbf{C}\hat{\mathbf{D}} \right\Vert_F} = \frac{\sqrt{P}\hat{\mathbf{D}}}{\left\Vert \text{diag}\left(e^{j\boldsymbol{\theta}_\mathbf{A}}\right)\mathbf{C}\hat{\mathbf{D}} \right\Vert_F}
\end{equation}

\begin{figure*}[t]
    \centering
    \includegraphics[width=0.86\linewidth]{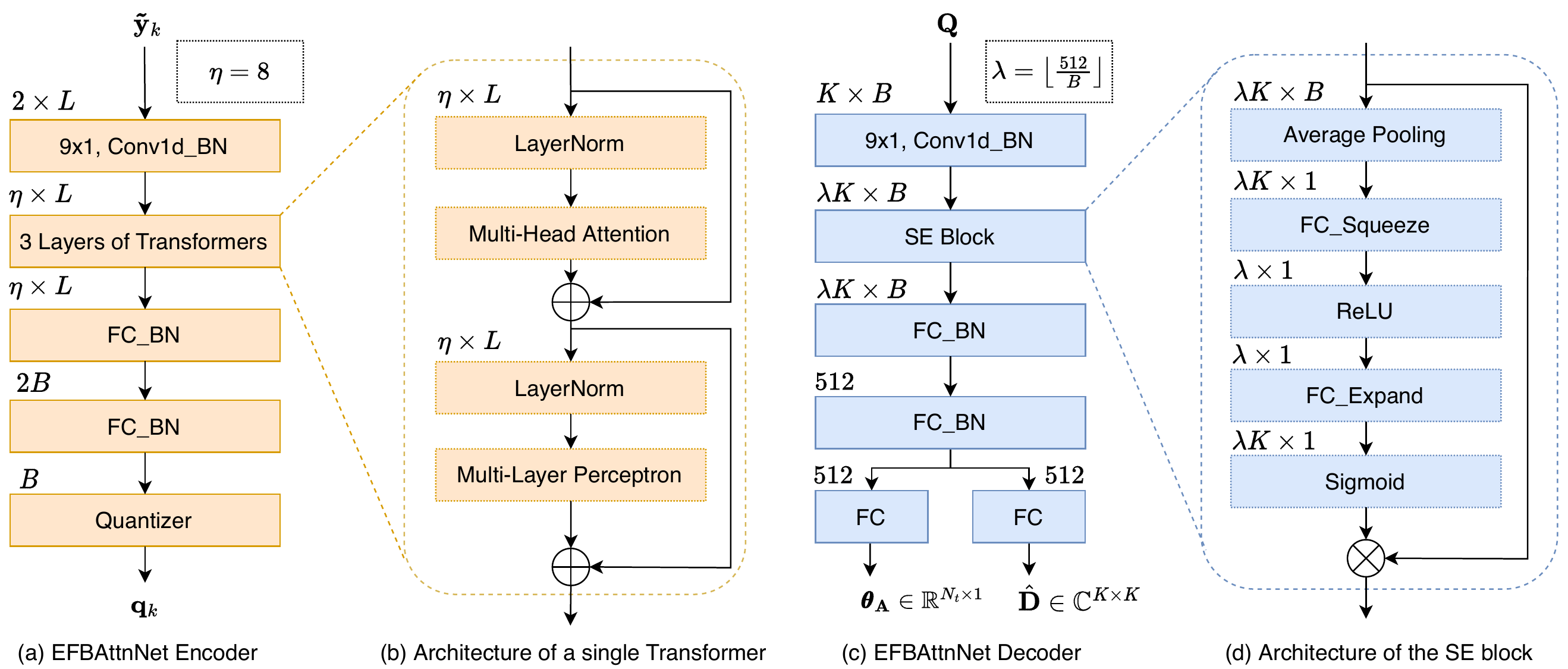}
    \caption{The design of the proposed EFBAttnNet. The input dimension is marked on top of each module with the batch size omitted.}
    \label{ImageEFBAttnNet}
\end{figure*}

Based on equation (\ref{EquationReceivedSignal}) and the beamformers designed above, the systematic beamforming sum rate can be given by adding up the achievable rates of all users as follows.
\begin{equation}
    R = \sum_{k=1}^K\log_2\left(1 + \frac{\left\vert\mathbf{h}_k^H\mathbf{A}\mathbf{C}\mathbf{d}_k\right\vert^2}{\sum_{i=1,i\neq k}^K\left\vert\mathbf{h}_k^H\mathbf{A}\mathbf{C}\mathbf{d}_i\right\vert^2 + \sigma^2}\right)
\end{equation}

\section{The Proposed EFBAttnNet} \label{SectionEFBAttnNet}

In this section, the proposed EFBAttnNet will be introduced in detail. The architecture of the EFBAttnNet encoder is depicted in Fig. \ref{ImageEFBAttnNet} (a). The real and imaginary part of $\mathbf{\tilde{y}}_k$ are concatenated as two channels and a $9\times 1$ convolutional layer expands the channels from $2$ to $8$ for feature enrichment.

Three serial transformer \cite{ashish2017attention} layers serve as the encoder attention module, and the architecture of each transformer is given in Fig. \ref{ImageEFBAttnNet} (b). The self-attention structure is good at learning from inputs with complicated inner-relationship, like estimating channel from the pilot signal $\mathbf{\tilde{y}}_k$.

After the attention modules, a two-layer inverted pyramid FC network with batch normalization (BN) is applied to adjust the output dimension. A final quantizer is responsible for the feature binarization since the uplink feedback requires a bitstream. We adopt the simple sign quantizer as (\ref{EquationQuantizer}) and a sigmoid-adjusted straight-through estimator is used during training to allow the gradient backpropagation of the non-differentiable sign function following \cite{Sohrabi2021deep}.
\begin{equation} \label{EquationQuantizer}
    \mathcal{Q}(x) = \text{Sign}(\text{Sigmoid}(x) - 0.5) \in \{+1, -1\},
\end{equation}
where $\mathcal{Q}(\cdot)$ stands for the quantizer.

Notably, each user has an independent EFBAttnNet encoder network while there is only one decoder network at the BS as it is demonstrated in Fig. \ref{ImageSystemModel}. The input of the EFBAttnNet decoder is also first expanded by a 1D convolutional layer with kernel size $9\times 1$ as we can see in Fig. \ref{ImageEFBAttnNet} (c). The expansion factor $\lambda$ is set to $\left\lfloor \frac{512}{B} \right\rfloor$ to stablize the decoder complexity under different number of the feedback bits $B$.

A squeeze and excitation (SE) \cite{jie2020squeeze} block is added after the convolutional expander to offer channel-wise attention. As shown in Fig. \ref{ImageEFBAttnNet} (d), an attention multiplier is learned to recalibrate the channel weight, which benefits the feature extraction of the two following FC layers. Finally, two extra FC layers are placed at the end to generate the targeted beamformers with the correct dimensions. Notably, the ReLU activation function is added after each BN layer, which is left out in Fig. \ref{ImageEFBAttnNet} for simplicity.

The main idea of the EFBAttnNet design is based on the attention mechanism. With the help of the transformer and the SE block, the network can largely reduce its complexity while maintaining the CSI feature extraction capability. Moreover, an unbalanced structure is specially designed so that the complexity of the encoder at the resource-sensitive UE is much lower than that of the decoder. Overall, the proposed EFBAttnNet is designed for more practical DL-based feedback and beamforming.

\section{Numerical Results} \label{SectionNumericalResults}
\subsection{Experimental Settings and Benchmarks}
For system settings, the number of paths in the SV channel model $L_\text{path}$ is set to $2$ while the transmitting power $P$ is set to $1$. The signal-to-noise ratio (SNR) of $10$ dB is used for all the experiments. For DL-related settings, we train the proposed EFBAttnNet in an unsupervised way that the loss function is set as the opposite sum rate $-R$. Adam optimizer and cosine annealing learning rate (LR) scheduler with initial LR $10^{-3}$ are applied following \cite{lu2020multiresolution} with a batch size of $1000$. Based on the channel model in (\ref{EquationSV}), the network is trained for $300$ epochs with $200$ batches in each epoch. All the sum rate performances are tested on $10000$ independently generated channels for a fair comparison.

To show the effectiveness of the proposed EFBAttnNet, the following methods are used as benchmarks:

\subsubsection{HP-sub/Full CSI} Ideal estimation and feedback are assumed and the BS has the full CSI. A zero forcing-based traditional beamforming method named HP-sub \cite{Guo2017subarray} is applied.
\subsubsection{HP-sub/OMP-CE/Infinite Feedback}
The UEs obtain the downlink CSI through channel estimation using the orthogonal matching pursuit (OMP) algorithm. Then the HP-sub beamforming is applied at the BS with ideal feedback.
\subsubsection{HP-sub/OMP-CE/Finite Feedback}
After the CSI is estimated at UEs with OMP, each UE quantizes the channel parameters $\{\alpha_{ik}, \theta_{ik} \}_{i = 1}^{L_\text{path}}$ into $B$ bits for limited feedback.
\subsubsection{EFBRefineNet} The previous SOTA network structure in \cite{Wu2022deep} is reimplemented to our subarray EFB pipeline for fair performance comparison.

\begin{figure}[!t]
    \centering
    \includegraphics[width=0.85\linewidth]{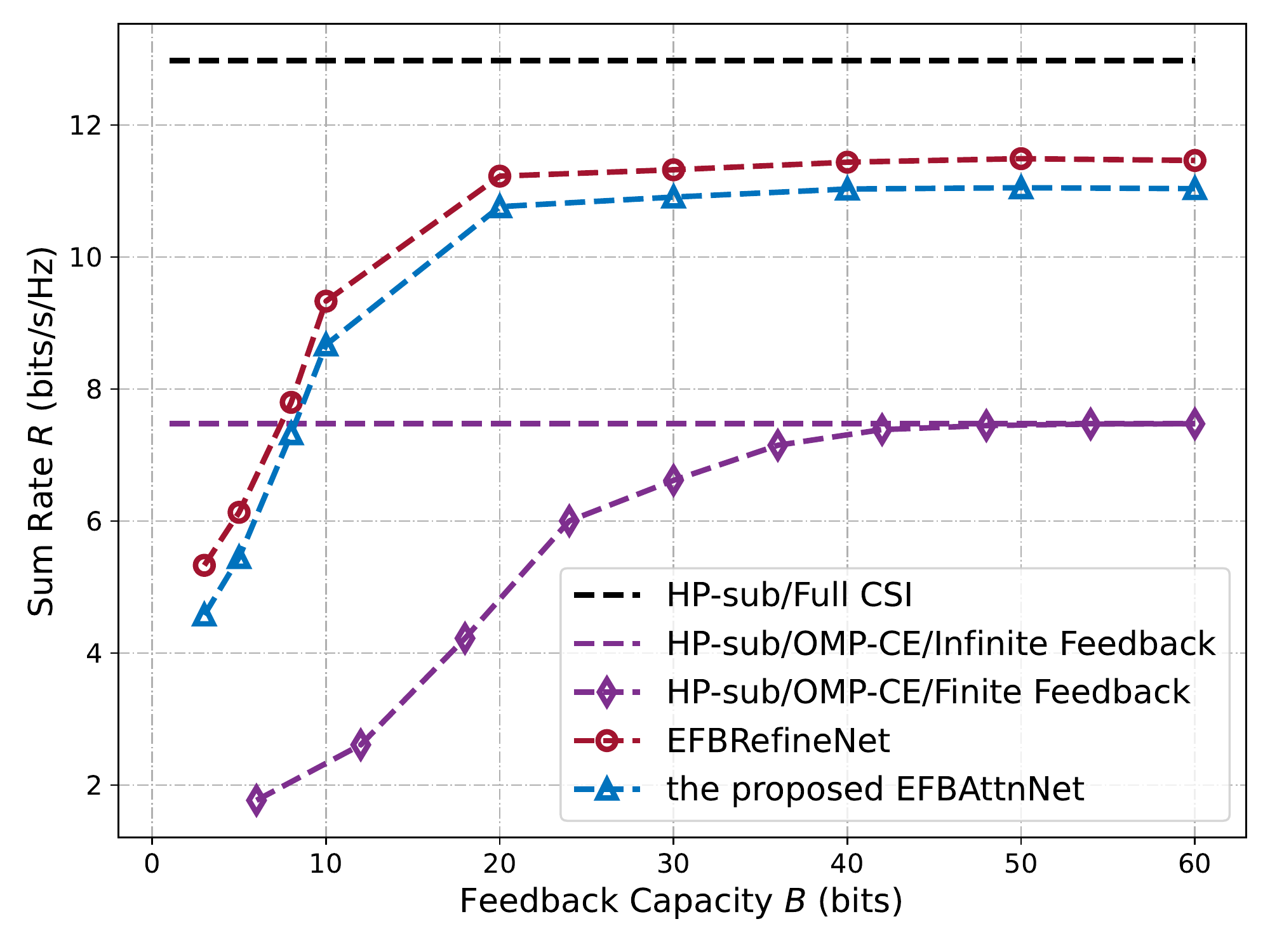}
    \caption{Sum rate performance of different methods in a two-user subarray hybrid beamforming FDD MIMO system with $N_t=64$, $L=8$.}
    \label{ImagePerformanceK2M64L8}
\end{figure}

\begin{table}[!t]
\caption{Complexity Comparison under Different $B$ with $K=2, L=8$. \label{TableComplexity}}
\centering
\resizebox{\linewidth}{!}{
    \begin{tabular}{ c c | r r | r r }
    \Xhline{0.8pt}
    \multirow{2}{*}{\textbf{B}} & \multirow{2}{*}{\textbf{Methods}}
        & \multicolumn{2}{c|}{\textbf{Params}} & \multicolumn{2}{c}{\textbf{FLOPs}} \\
        & & Total$^\mathrm{a}$ & UEs & Total & UEs\\
    \Xhline{0.8pt}
    \multirow{2}{*}{10}
        & EFBRefineNet\cite{Wu2022deep} & 4376K & 1103K & 4453K & 1117K \\
        & EFBAttnNet & \textbf{849K} & \textbf{12K} & \textbf{1021K} & \textbf{100K} \\
    \hline
    \multirow{2}{*}{40}
        & EFBRefineNet\cite{Wu2022deep} & 4530K & 1134K & 4607K & 1148K \\
        & EFBAttnNet & \textbf{821K} & \textbf{26K} & \textbf{994K} & \textbf{114K} \\
    \Xhline{0.8pt}
    \multicolumn{6}{l}{$^{\mathrm{a}}$ The total complexity of decoder and the two encoders.} \\
    \end{tabular}}
\end{table}

\subsection{Beamforming Performance and Complexity Analysis}

In this section, the sum rate performance of the proposed EFBAttnNet is compared with the aforementioned benchmarks. A two-user subarray EFB system is considered with $N_t=64$, which means that each RF chain at the BS is connected to $32$ antennas as explained in (\ref{EquationMergeBeamformer}) and (\ref{EquationC}).

\begin{figure}[!t]
    \centering
    \includegraphics[width=0.85\linewidth]{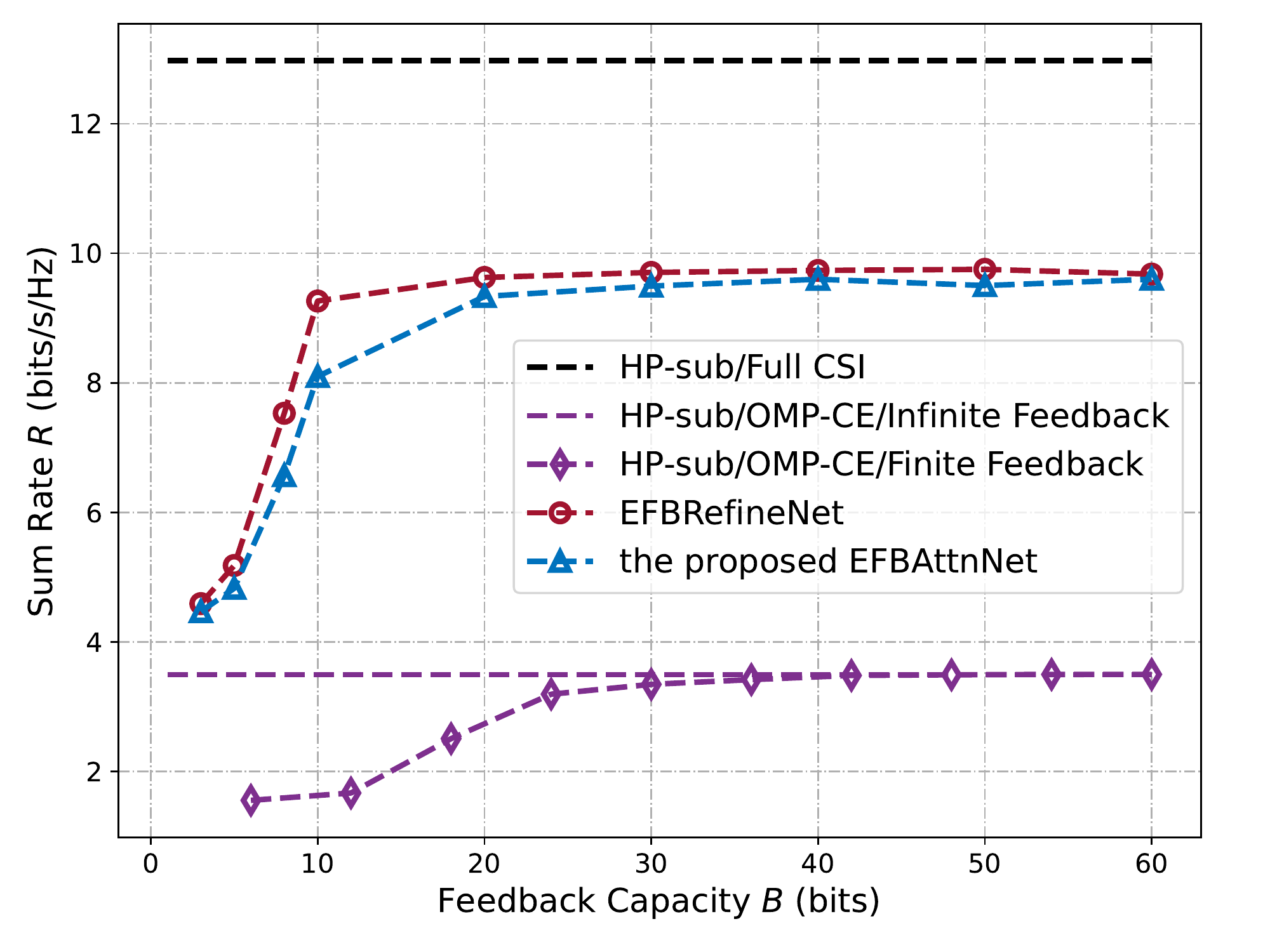}
    \caption{Sum rate performance of different methods in a two-user subarray hybrid beamforming FDD MIMO system with $N_t=64$, $L=4$.}
    \label{ImagePerformanceK2M64L4}
\end{figure}

As we can see from Fig. \ref{ImagePerformanceK2M64L8}, the proposed EFBAttnNet requires only $10$ feedback bits to outperform the traditional benchmark with ideal feedback when $L=8$. Moreover, the performance loss of the EFBAttnNet is much smaller than that of the traditional benchmark when the number of pilots further reduces from $8$ to $4$ as depicted in Fig. \ref{ImagePerformanceK2M64L4}. This proves that the proposed method can extract channel features and learn effective subarray hybrid beamformers from extremely limited pilots ($L \ll N_t$), which is very hard for the traditional block-based EFB pipeline.

When compared with the previous SOTA EFBRefineNet, the proposed EFBAttnNet trades a small performance loss for a huge decrease in complexity. As it is shown in Table \ref{TableComplexity}, the parameter size and the floating point operations (FLOPs) are largely reduced especially for the encoder at the resource-sensitive UEs. Moreover, the complexity ratio of a single encoder and decoder decreases from $16\%$ to around $6\%$. Such an unbalanced structure is also good for practical deployment.

\section{Conclusion} \label{SectionConclusion}
In this paper, a novel channel estimation, feedback, and beamforming network named EFBAttnNet was proposed for FDD massive MU-MIMO systems. In order to improve the practicality of deployment, the subarray hybrid beamforming architecture was applied for better energy efficiency. Additionally, the attention mechanism was used to extract better channel features at a low cost. Experiments proved the superiority of the proposed EFBAttnNet over the traditional benchmarks. Further comparison with the SOTA DL benchmark showed that the EFBAttnNet largely reduced the complexity especially for the resource-sensitive UEs with only a small sum rate loss.

\ifCLASSOPTIONcaptionsoff
  \newpage
\fi

\bibliographystyle{IEEEtran}
\bibliography{EFBAttnNet.bib}

\begin{thebibliography}{10}
\providecommand{\url}[1]{#1}
\csname url@samestyle\endcsname
\providecommand{\newblock}{\relax}
\providecommand{\bibinfo}[2]{#2}
\providecommand{\BIBentrySTDinterwordspacing}{\spaceskip=0pt\relax}
\providecommand{\BIBentryALTinterwordstretchfactor}{4}
\providecommand{\BIBentryALTinterwordspacing}{\spaceskip=\fontdimen2\font plus
\BIBentryALTinterwordstretchfactor\fontdimen3\font minus
  \fontdimen4\font\relax}
\providecommand{\BIBforeignlanguage}[2]{{%
\expandafter\ifx\csname l@#1\endcsname\relax
\typeout{** WARNING: IEEEtran.bst: No hyphenation pattern has been}%
\typeout{** loaded for the language `#1'. Using the pattern for}%
\typeout{** the default language instead.}%
\else
\language=\csname l@#1\endcsname
\fi
#2}}
\providecommand{\BIBdecl}{\relax}
\BIBdecl

\bibitem{Molisch2017hybrid}
A.~F. Molisch, V.~V. Ratnam, S.~Han, Z.~Li, S.~L.~H. Nguyen, L.~Li, and
  K.~Haneda, ``Hybrid beamforming for massive {MIMO}: A survey,'' \emph{IEEE
  Communications Magazine}, vol.~55, no.~9, pp. 134--141, 2017.

\bibitem{wen2018deep}
C.-K. Wen, W.-T. Shih, and S.~Jin, ``Deep learning for massive {MIMO} {CSI}
  feedback,'' \emph{IEEE Wireless Communications Letters}, vol.~7, no.~5, pp.
  748--751, 2018.

\bibitem{lu2020multiresolution}
Z.~Lu, J.~Wang, and J.~Song, ``Multi-resolution {CSI} feedback with deep
  learning in massive {MIMO} system,'' in \emph{ICC 2020 - 2020 IEEE
  International Conference on Communications (ICC)}, 2020, pp. 1--6.

\bibitem{Ma2021model}
X.~Ma, Z.~Gao, F.~Gao, and M.~Di~Renzo, ``Model-driven deep learning based
  channel estimation and feedback for millimeter-wave massive hybrid {MIMO}
  systems,'' \emph{IEEE Journal on Selected Areas in Communications}, vol.~39,
  no.~8, pp. 2388--2406, 2021.

\bibitem{guo2021deep-FB-BF}
J.~Guo, C.-K. Wen, and S.~Jin, ``Deep learning-based {CSI} feedback for
  beamforming in single- and multi-cell massive {MIMO} systems,'' \emph{IEEE
  Journal on Selected Areas in Communications}, vol.~39, no.~7, pp. 1872--1884,
  2021.

\bibitem{Sohrabi2021deep}
F.~Sohrabi, K.~M. Attiah, and W.~Yu, ``Deep learning for distributed channel
  feedback and multiuser precoding in {FDD} massive {MIMO},'' \emph{IEEE
  Transactions on Wireless Communications}, vol.~20, no.~7, pp. 4044--4057,
  2021.

\bibitem{Gao2022data}
Z.~Gao, M.~Wu, C.~Hu, F.~Gao, G.~Wen, D.~Zheng, and J.~Zhang, ``Data-driven
  deep learning based hybrid beamforming for aerial massive {MIMO}-{OFDM}
  systems with implicit {CSI},'' \emph{IEEE Journal on Selected Areas in
  Communications}, vol.~40, no.~10, pp. 2894--2913, 2022.

\bibitem{Wu2022deep}
M.~Wu, Z.~Gao, Z.~Gao, D.~Wu, Y.~Yang, and Y.~Huang, ``Deep learning-based
  hybrid precoding for {FDD} massive {MIMO}-{OFDM} systems with a limited pilot
  and feedback overhead,'' in \emph{2022 IEEE International Conference on
  Communications Workshops (ICC Workshops)}, 2022, pp. 318--323.

\bibitem{ashish2017attention}
A.~Vaswani, N.~Shazeer, N.~Parmar, J.~Uszkoreit, L.~Jones, A.~N. Gomez,
  L.~Kaiser, and I.~Polosukhin, ``Attention is all you need,'' in
  \emph{Proceedings of the 31st International Conference on Neural Information
  Processing Systems}, 2017, pp. 6000--6010.

\bibitem{jie2020squeeze}
J.~Hu, L.~Shen, S.~Albanie, G.~Sun, and E.~Wu, ``Squeeze-and-excitation
  networks,'' \emph{IEEE Transactions on Pattern Analysis and Machine
  Intelligence}, vol.~42, no.~8, pp. 2011--2023, 2020.

\bibitem{Guo2017subarray}
Y.~Guo, L.~Li, X.~Wen, W.~Chen, and Z.~Han, ``Sub-array based hybrid precoding
  design for downlink millimeter-wave multi-user massive {MIMO} systems,'' in
  \emph{2017 9th International Conference on Wireless Communications and Signal
  Processing (WCSP)}, 2017, pp. 1--4.

\end{thebibliography}



\end{document}